\documentstyle[pra,aps]{revtex}
\begin{document}


\title{\bf ROBUST ENTANGLEMENT IN ATOMIC SYSTEMS \\ VIA
$\Lambda$-TYPE PROCESSES}

\author{\"{O}zgur Cak{\i}r, M. Ali Can, Alexander Klyachko, and
Alexander Shumovsky}

\address{Faculty of Science, Bilkent University, Bilkent, Ankara,
06533 Turkey}

\maketitle

\begin{abstract}
It is shown that the system of two three-level atoms in $\Lambda$
configuration in a cavity can evolve to a long-lived maximum
entangled state if the Stokes photons vanish from the cavity by
means of either leakage or damping. The difference in evolution
picture corresponding to the general model and effective model
with two-photon process in two-level system is discussed.
\end{abstract}


\newpage

\section{Introduction}

During the last decade, the problem of engineered entanglement in
atomic systems has attracted a great deal of interest (see
\cite{1,2,3,4,5,6,7,8} and references therein). In particular, the
atomic entangled states were successfully realized through the use
of cavity QED \cite{1} and technique of ion traps \cite{3}. At
present, one of the most important problems under consideration is
how to make a long-lived and easy-monitored atomic entangled state
with existing experimental technique.

An interesting scheme has been proposed recently \cite{9}. In this
scheme, the two identical atoms are placed into a cavity tuned to
resonance with one of the dipole-allowed transitions. Initially
both atoms are prepared in the ground state, while the cavity
field consists of a single photon. It is easy to show that the
atom-field interaction leads in this case to a maximum atomic
entangled state such that the single excitation is shared between
the two atoms with equal probability. It was proposed in \cite{9}
to consider the absence of photon leakage from a nonideal cavity
as a signal that the atomic entangled state has been created. The
scheme can also be generalized on the case of any even number of
atoms $2n$, sharing $n$ excitations. In this case, the atomic
entangled states are represented by the so-called $SU(2)$ phase
states \cite{10}.

Another interesting proposal is to use a strong coherent drive to
provide the multipartite entanglement in a system of two-level
atoms in a high-Q cavity \cite{11}. This approach can be used to
produce the atomic entanglement as well as that of atoms and
cavity modes and even of different cavity modes.

In the schemes of Refs. \cite{9,11}, the lifetime of the
entanglement is defined by the specific time scale of the
dipole-allowed radiative processes in atoms. Unfortunately this
lifetime is usually quite short \cite{12}.

Generally speaking, the lifetime of atomic entanglement is
specified by the interaction of atoms with environment. For
example, in the model of Ref. \cite{10} the environment is
represented by the vacuum field that causes emission of a photon
getting out of the cavity.

The interaction with environment can also be used to create a
long-lived entanglement in atomic systems. For example, the
initially non-entangled system may evolve to an entangled state
connected with the atomic states that cannot be depopulated by
radiation decay. In this case, the lifetime of the entangled state
is specified by the considerably long nonradiative processes.
Possible realization is provided by the use of three-level
$\Lambda$-type process instead of the two-level scheme of Refs.
\cite{9,11}. The process is illustrated by Fig. 1. Here the levels
$1$ and $2$ are connected by the electric dipole transitions as
well as the levels $2$ and $3$. In turn, the dipole transition
between the levels $3$ and $1$ is forbidden because of the parity
conservation \cite{13}. The absorption of pumping photon by the
transition $1 \leftrightarrow 2$ with further jump of the electron
to the level $3$ accompanied by emission of "Stokes" photon can be
interpreted as a kind of Raman process in atomic system with
emission of Stokes photon (see \cite{14} and references therein).

It is clear that the atom excited to the level $3$ can change the
state either by absorption of the Stokes photon resonant with
respect to the transition $3 \leftrightarrow 2$ or trough a
nonradiative decay. Assume now that the two identical
$\Lambda$-type atoms are placed into a cavity of high quality with
respect to the pumping photons resonant to the transition $1
\leftrightarrow 2$. We also assume that the Stokes photons created
by the transition $2 \rightarrow 3$  either leave the cavity
freely or are absorbed by the cavity walls. Then, the atom-field
interaction may lead to creation of maximum entangled atomic state
\begin{eqnarray}
\frac{1}{\sqrt{2}} (|3,1 \rangle + |1,3 \rangle ), \label{1}
\end{eqnarray}
whose lifetime is determined by the slow processes of nonradiative
decay.

The above scheme has been proposed in Ref. \cite{10} and briefly
discussed in Ref. \cite{15}. The main objective of present paper
is to consider in details the evolution towards the long-lived
atomic entangled state (1) and feasibility of the scheme with
present experimental technique.

The paper is organized as follows. In Sec. II we discuss the model
Hamiltonians that can be used to describe the process under
consideration. Viz, we discuss the model of a one-photon
three-level interaction  and  an effective model of two-photon
process in two-level system. Then, in Sec. III, we examine the
irreversible dynamics in a cavity with leakage of Stokes photons,
leading to the state (1). We show that both models describe the
exponential evolution to the state (1). At the same time, the
effective model that corresponds to a rough time scale is unable
to take into account the possible oscillations of population
between the states $1$ and $2$. Let us stress that the monitoring
of Stokes photons outside the cavity can be used to detect the
atomic entangled state (1) in the case under consideration.

Another way of creation of state (1) through the use of a cavity
with very low quality with respect to the Stokes photons is
discussed in Sec. IV. Finally, in Sec. V, we discuss the possible
realization of the scheme under consideration.

\section{The models of $\Lambda$-type process}

Assume that a system of $N$ identical three-level atoms with
$\Lambda$-type transitions shown in Fig. 1 interacts with the
cavity mode close to resonance with $1 \leftrightarrow 2$
transition and with the Stokes radiation that can leave the cavity
freely. Then, following \cite{13,14}, we can choose the model
Hamiltonian in the following form
\begin{eqnarray}
H=H_0+H_{int}, \nonumber \\ H_0= \omega_P a^+_Pa_P+ \sum_k
\omega_{Sk} a^+_{Sk}a_{Sk} \nonumber \\ + \sum_f [ \omega_{21}
R_{22}(f) + \omega_{31} R_{33}(f)], \label{2} \\ H_{int}= \sum_f
\lambda_P R_{21}(f)a_P+ \sum_{f,k} \lambda_{Sk} R_{23}(f)a_{Sk}
+H.c. \label{3}
\end{eqnarray}
Here $a_P$ denotes the photon operator of the cavity mode with
frequency $\omega_P$, $a_{Sk}$ describes annihilation of Stokes
photon with frequency $\omega_{Sk}$, and $\omega_{21}$ and
$\omega_{31}$ are the frequencies of the corresponding atomic
levels with respect to the ground level $1$. The operator
\begin{eqnarray}
R_{ij}(f) =|i_f \rangle \langle j_f| \nonumber
\end{eqnarray}
gives the transition from level $j$ to level $i$, index $f$ marks
the number of atom. In Eq. (3), $\lambda_P$ and $\lambda_{Sk}$ are
the coupling constants, specifying the dipole transitions $2
\leftrightarrow 1$ and $3 \leftrightarrow 2$, respectively.
Summation over $k$ in (3) implies that the Stokes photons do not
feel the presence of cavity walls.  This summation involves the
modes distributed with a certain density in a narrow band near the
atomic transition frequency
\begin{eqnarray}
\omega_S \equiv \omega_{23}= \omega_{21} - \omega_{31}, \label{4}
\end{eqnarray}
corresponding to the natural line breadth.

Apart from the total electron occupation number, the model (2),
(3) has the two integrals of motion
\begin{eqnarray}
N_P=a^+_Pa_P + \sum_f \{ R_{22}(f) +R_{33}(f)\}\nonumber\\
N_S=\sum_ka^+_{Sk}a_{Sk}-\sum_fR_{33}(f).  \label{5}
\end{eqnarray}
Consider the system of only two atoms. Assume that both atoms are
prepared initially in the ground states $1$, the cavity contains a
single photon of frequency $\omega_P$, and Stokes field is in the
vacuum state. Then, because of the integrals of motion (5), the
evolution of the system occurs in a single-excitation domain of
the Hilbert space spanned by the vectors
\begin{eqnarray}
\left\{ \begin{array}{ll} | \psi_1 \rangle & =  |1,1 \rangle
\otimes |1_P \rangle \otimes |0_S \rangle  \\ | \psi^{(\pm)}_2
\rangle & =  \frac{1}{\sqrt{2}} (|1,2 \rangle \pm |2,1 \rangle )
\otimes |0_P \rangle \otimes |0_S \rangle  \\ | \psi_{3k}^{(\pm)}
\rangle & =  \frac{1}{\sqrt{2}} (|1,3 \rangle \pm |3,1 \rangle )
\otimes |0_P \rangle \otimes |1_{Sk} \rangle \end{array} \right.
\label{6}
\end{eqnarray}
By construction, the four states (6) labeled by the superscripts
$\pm$ manifest the maximum entanglement. No one of the states (6)
is the eigenstate of the Hamiltonian (2) and (3). At the same
time, it is easily seen that the action of operator (3) cannot
transform the states
\begin{eqnarray}
\{ | \psi_1 \rangle , | \psi_2^{(+)} \rangle , | \psi_{3k}^{(+)}
\rangle \} \label{7}
\end{eqnarray}
into the states
\begin{eqnarray}
\{ | \psi_2^{(-)} \rangle , | \psi_{3k}^{(-)} \rangle \} \label{8}
\end{eqnarray}
and vice versa. Thus, the evolution of the system from the initial
state $| \psi_1 \rangle$ occurs in the subspace spanned by the
three vectors (7). In this case, the states (8) can be discarded
from the basis (6).

Instead of the one-photon three-level model described by the
Hamiltonian (2) and (3), an effective model of two-photon process
can also be used under a certain condition \cite{16,17}. Viz, if
the cavity is tuned consistent with two-photon energy
conservation, i.e.
\begin{eqnarray}
E_3-E_1= \omega_1-\omega_2, \nonumber
\end{eqnarray}
we are left with only one detuning parameter
\begin{eqnarray}
\Delta =E_1-E_2-\omega_1=E_2-E_3-\omega_S. \nonumber
\end{eqnarray}
Here $E_i$ denotes the energy of corresponding atomic level. Then,
it was shown in \cite{16} that under the condition
\begin{eqnarray}
\Delta \gg E_3-E_1 \nonumber
\end{eqnarray}
the dynamics of the system is governed by the effective
Hamiltonian of the form
\begin{eqnarray}
H^{eff}= \omega_Pa^+_Pa_P + \omega_S a^+_Sa_S + \sum_f \omega_{31}
R_{33}(f) \nonumber \\ + \sum_f \lambda
[R_{31}(f)a^+_Sa_P+a^+_Pa_sR_{13}(f)] , \label{9}
\end{eqnarray}
describing the two-level transition with simultaneous absorption
of pumping photon and creation of Stokes photon. Here $\lambda$ is
an effective coupling constant.

\section{Dynamics described by the Hamiltonian (2),(3)}

Under the assumption that there are only two three-level
$\Lambda$-type atoms in the cavity and that the system is
initially prepared in the state $| \psi_1 \rangle$ in (5), in view
of the results of previous section we should choose the
time-dependent wave function as follows
\begin{eqnarray}
| \Psi (t) \rangle = C_1| \psi_1 \rangle +C_2| \psi_2 \rangle +
\sum_k C_{3k}| \psi_{3k} \rangle , \label{10} \\ C_1(0)=1, \quad
C_2(0)=0, \quad \forall k \quad C_k(0)=0 , \label{11}
\end{eqnarray}
using the reduced basis (7). Here for simplicity we use the
notations $| \psi_2 \rangle \equiv | \psi_2^{(+)} \rangle$ and $|
\psi_{3k} \rangle \equiv | \psi_{3k}^{(+)} \rangle$. The
time-dependent Schr\"{o}dinger equation with the Hamiltonian (2)
and (3) then leads to the following set of equations for the
coefficients in (11)
\begin{eqnarray}
\left\{ \begin{array}{ll} i\dot{C}_1 & =  \omega_P C_1 + \lambda_P
\sqrt{2} C_2
\\ i \dot{C}_2 & =  \omega_{21} C_2+ \lambda_P \sqrt{2} C_1+
\sum_k \lambda_{Sk} C_{3k} \\ i\dot{C}_{3k} & =  ( \omega_{31}+
\omega_{Sk} )C_{3k}+ \lambda_{Sk} C_2 \end{array} \right.
\label{12}
\end{eqnarray}
To find solutions of (12), let us represent the last equation in
(12) in the form
\begin{eqnarray}
C_{3k}(t)=-i \lambda_{Sk} \int_0^t
C_2(\tau)e^{i(\omega_{31}+\omega_{Sk})(\tau -t)}d \tau .
\label{13}
\end{eqnarray}
Then, we should take the time derivative in both sides of the
first equation in (12) and substitute the second equation together
with integral representation (13). We get
\begin{eqnarray}
i\ddot{C}_1=(\omega_P+\omega_{21})\dot{C}_1+i(\omega_{21} \omega_P
-2 \lambda_P^2)C_1 \nonumber \\ - \sum_k \lambda_{Sk}^2 \int_0^t
(i\dot{C}_1- \omega_P C_1)e^{i(\omega_{31}+\omega_{Sk})(\tau -t)}
d \tau . \nonumber
\end{eqnarray}
Carrying out the integration by parts, we get the following
integro-differential equation with respect to only one unknown
variable $C_1(t)$:
\begin{eqnarray}
i\ddot{C}_1=( \omega_P+\omega_{21})\dot{C}_1+i( \omega_{21}
\omega_P-2 \lambda_P^2  \nonumber \\ - \sum_k \lambda_{Sk}^2 )C_1+
i \sum_k \lambda_{Sk}^2 e^{-i(\omega_{31}+\omega_{Sk})t} - \sum_k
\lambda_{Sk}^2 (\omega_{31}+\omega_{Sk} \nonumber \\ -\omega_P)
\int_0^t C_1(\tau)e^{i(\omega_{31}+\omega_{Sk})(\tau -t)} d \tau
\nonumber
\\ \label{14}
\end{eqnarray}
This integro-differential equation can be analyzed through the use
of Laplace transformation, which is usually considered in the
context of natural line breadth (see Ref. \cite{19}). Let us
emphasize that, in contrast to the conventional Wigner-Weisskopf
theory, Eq. (14) contains the second-order derivative in addition
to the first-order term. In other words, it should describe
oscillations together with the decay. We have
\begin{eqnarray}
\int_0^{\infty} C_1(t)e^{-st}dt = {\cal L}(C_1), \nonumber \\
\int_0^{\infty} \dot{C}_1(t)e^{-st}dt=s{\cal L}-1, \nonumber \\
\int_0^{\infty} \ddot{C}_1(t)e^{-st}dt=s^2{\cal L}-s-\dot{C}_1(0)
= s^2{\cal L}-s+i \omega_P. \nonumber
\end{eqnarray}
Then, Eq. (14) is reduced into the following algebraic equation
with respect to ${\cal L}$:
\begin{eqnarray}
{\cal L}[is^2-s( \omega_P+\omega_{21})  -i( \omega_{21} \omega_P
-2 \lambda_P^2 - \sum_k \lambda_{Sk}^2)] \nonumber \\ =-(
\omega_P+\omega_{21})+ i \sum_k
\lambda^2_{Sk}\frac{s+i(\omega_{31}+\omega_{Sk})}{s^2 +
(\omega_{31}+\omega_{Sk})^2} - \int_0^{\infty} e^{-st} \{ \sum_k
\lambda_{Sk}^2 \nonumber \\ \times ( \omega_{31}+ \omega_{Sk} -
\omega_P) \int_0^t C_1(\tau)e^{i(\omega_{31}+\omega_{Sk})(\tau
-t)}d \tau \} dt . \label{15}
\end{eqnarray}
The last term in the right-hand side of this expression can be
represented as follows
\begin{eqnarray}
\int_0^{\infty} e^{-st} \{ \int_0^t e^{i( \omega_{31} +
\omega_{Sk} )( \tau -t)}C_1( \tau )d \tau \}dt \nonumber \\ =
\int_0^{\infty} C_1( \tau )e^{i( \omega_{31} + \omega_{Sk}) \tau}d
\tau \int_{\tau}^{\infty} e^{-[s+i( \omega_{31} +
\omega_{Sk})]t}dt \nonumber \\ = \frac{\cal L}{s+i( \omega_{31} +
\omega_{Sk} )} . \nonumber
\end{eqnarray}
Thus, Eq. (15) takes the form
\begin{eqnarray}
{\cal L}= \left[ i \sum_k \lambda_{Sk}^2
\frac{s+i(\omega_{31}+\omega_{Sk})}{s^2 +
(\omega_{31}+\omega_{Sk})^2} -(\omega_P+\omega_{21}) \right]
\nonumber \\ \times  [ is^2-s(
\omega_P+\omega_{21})+i(\omega_P\omega_{21} -2 \lambda_P^2- \sum_k
\lambda_{Sk}^2 \nonumber \\ + \sum_k \frac{ \lambda_{Sk}^2
(\omega_{31}+\omega_{Sk}-\omega_P)}{s+i(\omega_{31}+\omega_{Sk})}
]^{-1} . \label{16}
\end{eqnarray}
Then, the exact form of the time behavior of the coefficient
$C_1(t)$ in Eq. (10) is governed by the inverse Laplace
transformation
\begin{eqnarray}
C_1(t)= \frac{1}{2 \pi i} \int_{\epsilon -i \infty}^{\epsilon +i
\infty} e^{st}{\cal L}(s)ds , \label{17}
\end{eqnarray}
where $\epsilon$ is a small real positive number and $s$ in the
integrand is considered as a complex argument. As soon as the
explicit time behavior of $C_1(t)$ is known, the other
coefficients in (10) can be defined through the use of Eqs. (12)
and (13).

In particular, it follows from Eqs. (6), (10), and (11) that the
probability to have the atomic entangled state (1) has the form
\begin{eqnarray}
\sum_k |C_{3k}|^2=1-|C_1(t)|^2-|C_2(t)|^2 \nonumber \\
=1-|C_1(t)|^2- \frac{|i\dot{C}_1(t)- \omega_P C_1(t)|^2}{2
\lambda_P^2} . \label{18}
\end{eqnarray}
Thus, the Eqs. (16) and (17) completely determine the probability
to have the robust entangled state (1).

It can be shown that Eq. (16) describes the reversible,
Poincar'{e} type behavior (e.g., see \cite{19}). The irreversible
evolution can be obtained under the further assumption that the
atomic transition $2 \leftrightarrow 3$ interacts with continuum
of Stokes modes rather than with a discrete spectrum. Similar
assumption is used within conventional Wigner-Weisskopf theory
\cite{18}. This means that we replace summation over $k$ by
integration over $\omega =ck$:
\begin{eqnarray}
\sum_k \cdots \rightarrow \int_{- \infty}^{\infty} \cdots \rho
(\omega) d \omega . \nonumber
\end{eqnarray}
Here the measure $\rho ( \omega )d \omega$ defines the density of
states of Stokes photons with different frequencies corresponding
to the natural line breadth. Unlike conventional Wigner-Weisskopf
theory, Eqs. (16) and (17) describe a superposition of exponential
decay and harmonic oscillations caused by the interaction between
the $1 \leftrightarrow 2$ transitions and cavity field.

Further analysis shows that, the coefficients $C_1$ and $C_2$ up
to the second order in $\lambda/(\Gamma-i\Delta_P)$ are
\begin{eqnarray}\label{19}
C_1(t)&\approx& \left
[-\frac{2\lambda^2}{(\Gamma-i\Delta)^2}e^{(-\Gamma+ i\Delta)t}
+(1+\frac{2\lambda^2}{(\Gamma-i\Delta)^2})e^{-\frac{2\lambda^2}
{\Gamma-i\Delta}t}\right ]e^{-i\omega_Pt}\nonumber\\
C_2(t)&\approx&-\frac{\sqrt{2}\lambda}{i\Gamma+\Delta}\left[e^{-\Gamma
t}-e^{-(\frac{2\lambda^2}{\Gamma-i\Delta}+i\Delta)t}\right]e^{-i\omega_{21}t}
\end{eqnarray}
where
\begin{eqnarray}
\Delta_P = \omega_P- \omega_{12} \nonumber
\end{eqnarray}
is the detuning factor for the pumping mode and
\begin{eqnarray}
\Gamma = \rho( \omega_S) \lambda_{Sk}(k= \omega_S/c) , \quad
\omega_S = \omega_{21}-\omega_{31}. \nonumber
\end{eqnarray}
Since $\rho(\omega_S)\gg 1$,  $\Gamma \gg \lambda_P ,
\lambda_{Sk}$ and the result (19) corresponds to the second order
order with respect to $\lambda_P/(\Gamma -i \Delta)$.

It is seen that Eq. (19) describes the damped oscillations of the
coefficient $C_1(t)$ in (10). According to (11), coefficient
$C_2(t)$ manifests similar behavior. Thus, the probability (18) to
get the robust entangled state tends to unit as $t \rightarrow
\infty$ (see Fig. 2). The decay time $\gamma^{-1}$ is defined by
the coupling constant and detuning parameter for the pumping mode
and by the width of the Stokes line specified by the parameter
$\Gamma$. Although Eq. (19) contains oscillating terms, the decay
is quite strong and the influence of oscillating terms cannot be
observed even when  $\Delta_P \leq \Gamma$. Only when detuning
strongly exceeds $\Gamma$, the oscillations start to contribute
significantly into the time evolution of probability (18).

Similar result can also be obtained in terms of the effective
Hamiltonian (9) \cite{15}. It should be stressed that the
assumptions made in the process of derivation of (9) lead to an
effective "roughing" of the time scale. The obliteration of the
level $2$ leads to the simultaneous neglect of the Rabi
oscillations between the levels $1$ and $2$. Therefore, the
effective model (9) is incapable of recognition of time
oscillations and gives only rough picture of purely exponential
evolution.

While the atomic system evolves to the maximum entangled state
(1), the Stokes photon leaves the cavity. Thus, the observation of
Stokes photon outside the cavity can be considered as a signal
that the robust entangled state has been prepared.

\section{Cavity with absorption of Stokes photons}

The atomic entangled state (1) can also be realized when the
Stokes mode is strongly damped in the cavity. For simplicity, we
still assume no damping for the pumping mode. At the same time,
the Stokes photons are supposed to be absorbed by the cavity
walls. In other words, we consider the case of high quality (with
respect to pumping field photons) and zero-temperature cavity,
corresponding to the most of the Ridberg maser experiments
\cite{20,21}. In this case, the effect of damping can be
calculated through the use of the so-called dressed-atom
approximation \cite{22}.

The model Hamiltonian, describing the process under consideration,
can be chosen as follows
\begin{eqnarray}
H=H_0+H_{int}, \nonumber \\ H_0= \omega_P a^+_Pa_P + \omega_S
a^+_Sa_S \nonumber \\ + \sum_f [\omega_{21} R_{22}(f)+ \omega_{31}
R_{33}(f)], \nonumber \\ H_{int}= \sum_f [\lambda_P R_{21}(f)a_P+
\lambda_S R_{23}(f)a_S]+H.c. \label{20}
\end{eqnarray}
This corresponds to the single-Stokes-mode approximation in (2)
and (3). The eigenstates of the Hamiltonian (20) have the form
\begin{eqnarray}
| \psi_0 \rangle = \frac{\lambda_S}{\epsilon} | \psi_1 \rangle -
\frac{\lambda_P \sqrt{2}}{\epsilon} | \psi_3 \rangle , \nonumber
\\
| \psi_{\pm} \rangle = \pm \frac{\lambda_P}{\epsilon} | \psi_1
\rangle + \frac{1}{\sqrt{2}} | \psi_2 \rangle \pm
\frac{\lambda_S}{\epsilon \sqrt{2}} | \psi_3 \rangle \label{21}
\end{eqnarray}
where $| \psi_1 \rangle$ coincides with the first state in (6), $|
\psi_2 \rangle = | \psi^{(+)}_2 \rangle$, and
\begin{eqnarray}
| \psi_3 \rangle = \frac{1}{\sqrt{2}} (|3,1 \rangle +|1,3 \rangle
) \otimes |0_P \rangle \otimes |1_S \rangle . \nonumber
\end{eqnarray}
In Eqs. (21),
\begin{eqnarray}
\epsilon = \sqrt{2 \lambda_P^2 + \lambda_S^2} . \nonumber
\end{eqnarray}
Under the assumption of exact resonance
\begin{eqnarray}
\omega_P = \omega_{21} = \omega_{31} + \omega_S \nonumber
\end{eqnarray}
that we use hereafter for simplicity, the corresponding
eigenvalues are
\begin{eqnarray}
H| \psi_0 \rangle = \omega_P | \psi_0 \rangle , \quad H|
\psi_{\pm} \rangle = ( \omega_P \pm \epsilon )| \psi_{\pm} \rangle
. \nonumber
\end{eqnarray}
Besides that, there is one more eigenstate
\begin{eqnarray}
| \psi_4 \rangle = \frac{1}{\sqrt{2}} (|3,1 \rangle +|1,3 \rangle
) \otimes |0_P \rangle \otimes |0_S \rangle , \label{22}
\end{eqnarray}
such that
\begin{eqnarray}
H| \psi_4 \rangle = \omega_{31} | \psi_4 \rangle . \nonumber
\end{eqnarray}
It is clear that this eigenstate corresponds to the maximum atomic
entanglement (1). Physically, this state is achieved when the
Stokes photon is absorbed by the cavity walls.

To take into account the cavity damping of Stokes photons,
consider interaction with a "phonon reservoir" responsible for the
absorption of photons by cavity walls \cite{18}. Then, the
Hamiltonian (20) should be supplemented with the term
\begin{eqnarray}
H_{loss}= \sum_q \eta_q (b^+_qa_S+a^+_Sb_q) + \sum_q \Omega_q
b^+_qb_q , \label{23}
\end{eqnarray}
where $b_q,b^+_q$ are the Bose operators of "phonons" in the
cavity walls.

The density matrix of the system can be chosen as follows
\begin{eqnarray}
\rho (t) = \sum_{j, \ell} \rho_{j \ell}(t) | \psi_j \rangle
\langle \psi_{\ell} |, \quad j, \ell = 0, \pm ,4, \label{24}
\end{eqnarray}
where $| \psi_j \rangle$ are the eigenstates (21) and (22) and
$\rho_{j \ell}(t)$ denote the time-dependent $c$-number
coefficients.

With the total Hamiltonian
\begin{eqnarray}
H_{tot}=H+H_{loss} \nonumber
\end{eqnarray}
in hand, we can now write the Master Equation, eliminating the
cavity degrees of freedom (e.g., see \cite{23}),
\begin{eqnarray}
\dot{\rho} =-i[H, \rho ]+ \kappa \{ 2a_S \rho a^+_S -a^+_Sa_S \rho
- \rho a^+_Sa_S \} , \label{25}
\end{eqnarray}
so that the contribution of (23) is taken into account effectively
through the Liouville term. Here $1/ \kappa$ is the lifetime of a
Stokes photon in the cavity and $Q =E_3 / \kappa$ is the quality
factor with respect to the Stokes photons ($E_3$ denotes the
energy of the level $3$). Let us choose the same initial condition
as in previous section, so that
\begin{eqnarray}
\rho(0) =| \psi_1 \rangle \langle \psi_1| \label{26}
\end{eqnarray}
where the initial state $| \psi_1 \rangle$ is expressed in terms
of the eigenstates (21) as follows
\begin{eqnarray}
|\psi_1 \rangle = \frac{\lambda_S}{\epsilon} | \psi_0 \rangle +
\frac{\lambda_P}{\epsilon} (| \psi_+ \rangle -| \psi_- \rangle ).
\nonumber
\end{eqnarray}
The equation (25) can now be solved numerically at different
values of parameter $\kappa$, specifying the absorption of Stokes
photons. The results are shown in Fig. 3. It is seen that the
system evolves to the robust atomic entangled state (1). The
stairs-like structure is again caused by competition between the
transitions $1 \leftrightarrow 2$ and $2 \leftrightarrow 3$.
Although such a behavior is an inherent property of the model
under consideration, the stairs become more visible with increase
of $\kappa$ (see Fig. 3).

Similar result can also be obtained within the framework of
effective model with the Hamiltonian (9) and damping described by
Eq. (23). In this case, the density matrix consists of only six
elements because the state $| \psi_0 \rangle$ in (21) should be
discarded and the states $| \psi_{\pm} \rangle$ are changed by the
states
\begin{eqnarray}
| \phi_{\pm} \rangle = \frac{1}{\sqrt{2}} (| \psi_1 \rangle \pm |
\psi_3 \rangle ) \nonumber
\end{eqnarray}
with the eigenvalues
\begin{eqnarray}
\varepsilon_{\pm} = \omega_P \pm \lambda \sqrt{2} . \nonumber
\end{eqnarray}
It should be stressed that the effective model does not show the
stairs-like behavior of $\rho_{44}(t)$.

\section{Summary and discussion}

In this paper, we have studied the quantum dynamics of a system of
two three-level atoms in the $\Lambda$ configuration interacting
with two modes of quantized electromagnetic field in a cavity
under the assumption that the Stokes-mode photons either leave the
cavity freely or are damped rapidly. It is shown that in both
cases the system evolves from the state when both atoms are in the
ground state and cavity contains a pumping photon into the robust
entangled state (1). The lifetime of this final state is defined
completely by the nonradiative processes and is therefore
relatively long.

In the case of cavity transparent for the Stokes photons, the
creation of Stokes photon signalizes the rise of atomic
entanglement. Such a photon can be monitored outside the cavity.

Let us stress that the general models with the Hamiltonians (2),
(3) and (20, that takes into account all three atomic levels,
admit a certain peculiarities in the evolution towards the robust
entangled state caused by the competition of transitions $1
\leftrightarrow 2$ and $2 \leftrightarrow 3$. The effective model
with adiabatically eliminated highest excited level is incapable
of description of these peculiarities, while predicts correct
asymptotic behavior. Moreover, the general model admits also a
number of intermediate maximum entangled states ($| \psi_2
\rangle$ and $| \psi_{3k} \rangle$ in Eq. (6)) that do not exist
in the effective model. Unfortunately, the lifetime of these
entangled states are defined by the dipole radiative processes and
are therefore too short.

It seems to be possible to realize the robust entangled state (1)
using the present experimental technique. Let us note in this
connection that the single-atom Raman process has been recently
observed \cite{24}.

We also note that the way of experimental preparation of
single-photon cavity excitation is well known \cite{21}. The atoms
can propagate through the cavity, using either the same opening or
two different openings. All measurements aimed at the detection of
atomic entanglement can be performed outside the cavity.

Although our results were obtained for a system of two atoms, they
can be generalized with easy on the case of big atomic clusters,
using the method of Ref. \cite{10}. In fact, it is possible to
show that a certain robust entanglement can be obtained in a
system with any even number $2N$ of three-level $\Lambda$-type
atoms initially prepared in the ground state and interacting with
$N$ pumping photons.

\section*{Acknowledgement}

One of the authors (A.Sh.) would like to thank Prof. J.H. Eberly,
Prof. A. Vourdas, and Prof. D.G. Welsch for many useful
discussions.

\begin{figure}
\caption{Scheme of the process and configuration of atomic levels
and transitions.}
\end{figure}

\begin{figure}
\caption{Time evolution of probability (18) to have the robust
entanglement at $\lambda_P=0.001\Gamma$ for (1)$\Delta_P=0$
(2)$\Delta_P=\Gamma$ (3)$\Delta_P=2\Gamma$ (4)$\Delta_P=4\Gamma$}
\end{figure}

\begin{figure}
\caption{Time evolution of $\rho_{44}(t)$ at $\lambda_P=\lambda_S$
and $\kappa = 0.01 \lambda_P$ (a) and $\kappa = 0.1 \lambda_P$
(b).}
\end{figure}

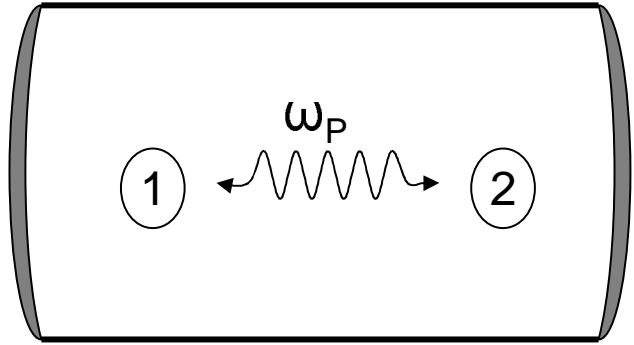
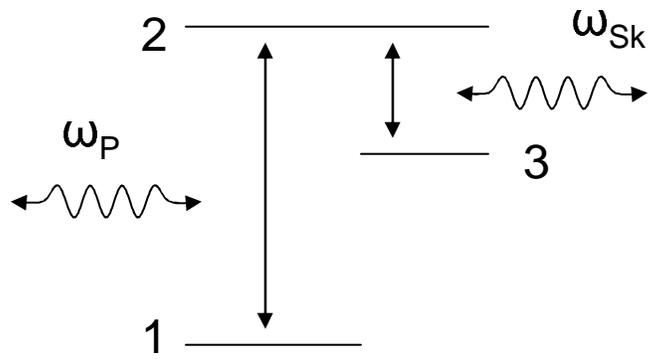
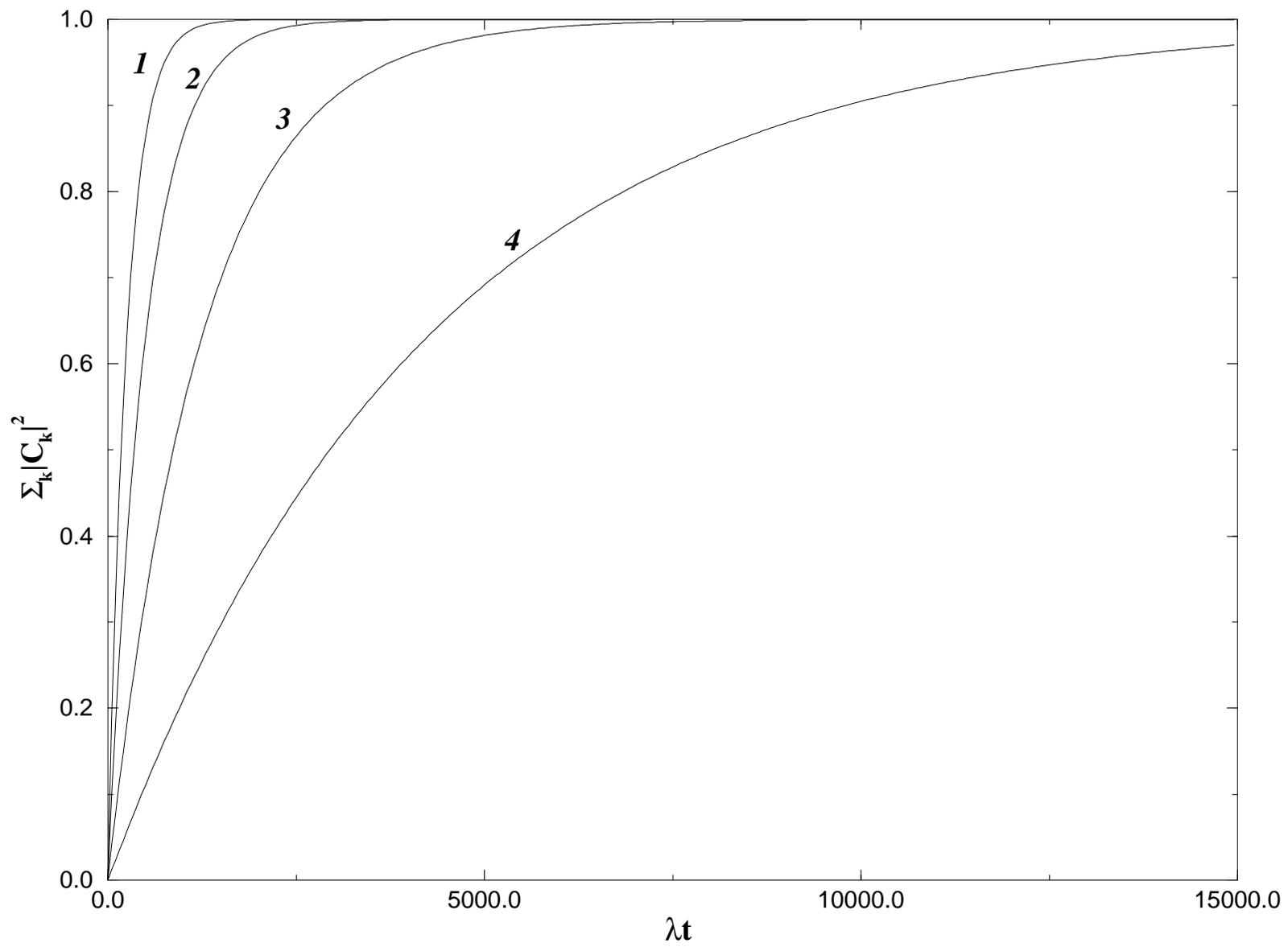
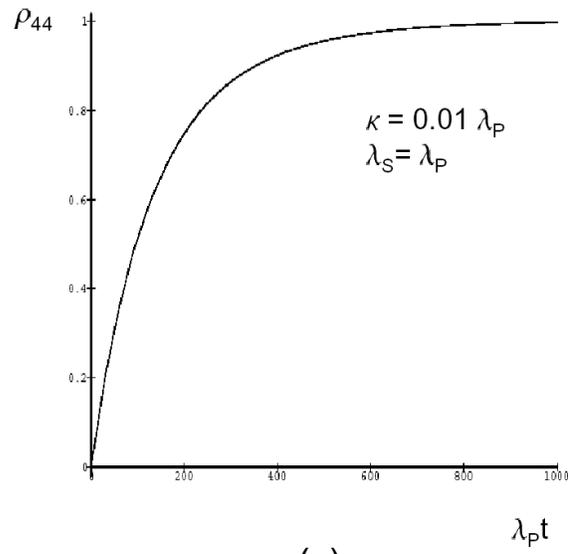
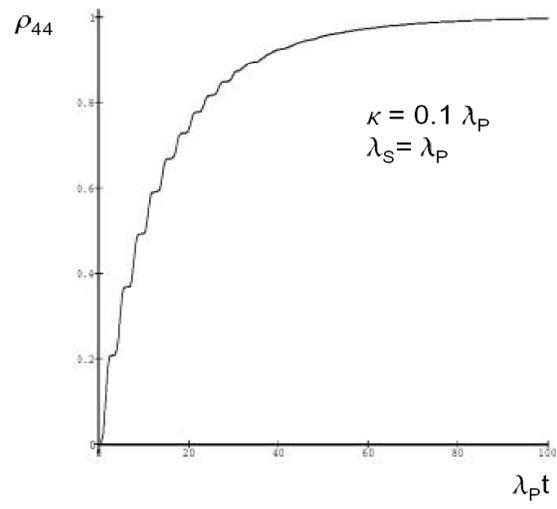
\end{document}